\documentclass[twocolumn,superscriptaddress,showpacs,prl,amsmath,amssymb]{revtex4}
\usepackage{epsfig}
\begin{document}

\title{
Noise spectra of stochastic pulse sequences: application to
large scale magnetization flips in the finite size 2D Ising model} 
\author{Zhi Chen and Clare C. Yu} 
\affiliation{Department of Physics and
Astronomy, University of California, Irvine, Irvine, California 92697}
\date{\today} 
\begin{abstract}
We provide a general scheme to predict and derive the contribution
to the noise spectrum of a
stochastic sequence of pulses from the distribution of pulse
parameters. An example is the magnetization noise spectra of a
2D Ising system near its phase transition. At $T\le T_c$, the low frequency
spectra is dominated by magnetization flips of nearly the entire system. We 
find that both the predicted and the analytically derived spectra
fit those produced from simulations. Subtracting this contribution
leaves the high frequency spectra which follow a power law set by 
the critical exponents.
\end{abstract}
\maketitle 

Noise due to random pulses is ubiquitous.
Examples include switching the rotational direction of
the flagellar motor in {\it Escherichia coli} bacteria
\cite{bacteria_Refs}, switching in electrical resistance
\cite{Weissman88}, and switching between
degenerate ordered phases of a finite size system.
Yet another example is crackling noise in
which slowly driven systems produce sudden discrete outbursts spanning
a broad range of sizes \cite{Sethna_nature01}. Instances of crackling
noise include the sound of paper crumpling, Barkhausen noise from
domain movement in ferromagnets~\cite{Celasco74,Zapperi_naturephy05},
and seismic activity during earthquakes \cite{Houston98}.

The question is how these pulses are reflected in the features of the 
power spectra commonly used to characterize noise. The answer could 
be used to estimate or predict the pulse noise spectrum as well as to
separate the pulse contribution to the noise spectrum from other sources.
For example, suppose one wants to determine the critical 
exponents of a second order phase transition from the noise spectra 
\cite{dAuriac82,CritExp,zhi_yuPRL07}. In a finite size system with a discrete 
broken symmetry, switching between degenerate ordered phases 
will also contribute to the noise spectra,
and it is important to separate out this contribution before
determining the critical exponents. 

\begin{figure}
\epsfysize=0.9\columnwidth{\rotatebox{-90}{\epsfbox{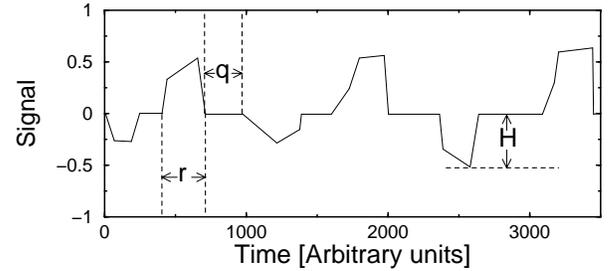}}}
\caption{A representative sequence of stochastic pulses. }
\label{simu_sig}
\end{figure}

Previous work calculated the noise from stochastic pulse sequences 
\cite{MachlupJAP54,Lukes61,Heiden69,CelascoJAP77}.  Machlup showed
that random telegraph noise consisting of square pulses with
exponentially distributed durations has a Lorentzian noise
spectrum \cite{MachlupJAP54}. Subsequently others
\cite{Lukes61,Heiden69,CelascoJAP77} considered a more general pulse
shape and distribution, though their theory cannot be applied if the
pulse shape depends on the pulse index $m$,
e.g., if the pulses alternate in sign as shown in Fig.~\ref{simu_sig}.
In this paper we determine the noise spectrum from a distribution of
pulse parameters for a much more general sequence of stochastic pulses
with amplitudes that can (but need not) depend on $m$.  
We then analyze the magnetization noise spectra
of a finite size 2D Ising system near the phase transition. We 
map the collective jumps (or flips) in the magnetization to a 
sequence of discrete pulses, and use our algorithm to predict
the contribution to the noise
using reasonable assumptions about the distribution of pulse
parameters and characteristic frequencies from the noise spectrum. 
We check this prediction with the spectra found from Monte Carlo simulations.
We then check our algorithm by determining the distribution of pulse
parameters from the magnetization time series, and 
derive noise spectra in excellent agreement with that
found from Monte Carlo simulations. We find that the low
frequency magnetization noise spectra have a distinct signature due to
the global magnetization flips of the whole spin system which is
particularly evident below $T_c$. Subtracting this contribution 
yields the high frequency power law noise
spectrum dictated by the critical exponents of the infinite system.

We now present a general formulation to find the ensemble averaged
noise spectrum of a sequence of stochastic pulses $X(t)$ consisting of
$K$ pulses.  As shown in Fig.~\ref{simu_sig}, each pulse has a duration
$r$ with an interval $q$ where the signal is zero between two
successive pulses. The maximum height $H(m)$ of the $m$th pulse has
two factors: $H(m)=h\cdot a(m)$, where $h$ is a random variable, and
$a(m)$ contains the functional dependence on $m$.  We only consider
the case where $a(m)=A^{m}$, $A$ is a constant, and $0<|A|\leq
1$. $|A|<1$ corresponds to a decay envelope. Denote $r$, $h$, and $q$
for the $m$th pulse by $r_m$, $h_m$, and $q_m$, respectively. $X(t)$
can be expressed as $X(t)=\sum_{m=1}^{K}x_m(t-t_m)$, where the $m$th
pulse is $x_m (t-t_m)$, and $t_m=\sum_{j=1}^{m-1} (r_j+q_j), m\ge2$ is
the corresponding time delay. (The first pulse starts at $t_1=0$.)
The mth pulse $x_m(t^{\prime})$ starts at $t^{\prime}=t-t_m=0$.

We denote the Fourier transform of the pulse $x_m(t^{\prime})$ by
$F_m(\omega,r_m,h_m,q_m)$. Then the Fourier transform $F_X(\omega)$ of
$X(t)=\sum_{m=1}^{K}x_m(t-t_m)$ is given by
\begin{equation}
F_X(\omega)=\sum_{m=1}^{K}F_m(\omega,r_m,h_m,q_m)e^{-i\omega t_{m}}.
\label{pow_fourier} 
\end{equation}
$\exp[-i\omega t_{m}]$ contains the phase relation between pulses.
The power spectrum of $X(t)$ is
$S_X(\omega)=2F_X^{*}(\omega)F_X(\omega)/\tau_K$, where $\omega >0$, and $\tau_K$
is the total duration of $X(t)$. $S_X(\omega)$ can be written as:
\begin{eqnarray}
&&\tau_K S_X(\omega)=2\sum_{m=1}^{K}|F_m(\omega,r_m,h_m,q_m)|^2
+4\mbox{Re}\bigg[\sum_{m=1}^{K-1}\sum_{n=1}^{K-m} \nonumber\\
&&F_n^*(\omega,r_n,h_n,q_n)F_{m+n}(\omega,r_{m+n},h_{m+n},q_{m+n})
e^{-i\omega (t_{m+n}-t_{n})}\bigg]
\label{pow_1} 
\end{eqnarray}

We assume the following: (a) The values of a pulse's parameters are
independent of those of other pulses except for $a(m)$.  (b) $r$, $q$
and $h$ have the combined distribution $p(r,h,q)$ which is same for
all pulses, and other pulse parameters (if any) are independent of
$r$, $h$ and $q$.  Then the ensemble averaged Fourier transform of the
pulse $x_m(t^{\prime})$ is $\int \int \int drdqdh
A^m\overline{F}(\omega,r,h,q) p(r,h,q)$, where $\omega$ is the angular
frequency, and the overline, e.g., $\overline{F}$, denotes the average of
over parameters other than $r$, $q$ and $h$
(e.g., different pulse shapes). (c) $\overline{F}(\omega,r,h,q)$ is
independent of the pulse index $m$.

From assumption (a), $F_m/A^m$ and $F_n/A^n$, as well as ($r_m+q_m$)
and ($r_n+q_n$) ($m\neq n$), are uncorrelated. After taking the
ensemble average over pulses and all parameters, we use assumptions
(b) and (c) to obtain:
\begin{eqnarray}
\langle \tau_KS_X(\omega)\rangle 
=
2BI_0
+4\mbox{Re}\left[AI_2I_3 R_{K}\right],
\label{pow_2} 
\end{eqnarray}
where $\langle ... \rangle$ is an ensemble average,
$B=\left(|A|^2-|A|^{2K+2}\right)/\left(1-|A|^2\right)$ 
for all $|A|^2<1$, and $B=K$ when 
$|A|^2=1$. We define
\begin{eqnarray}
R_{K}&=&\sum_{m=1}^{K-1}\sum_{n=1}^{K-m}(|A|^2)^n(AI_1)^{m-1}\nonumber\\
I_0(\omega)&=&\int_{-\infty}^{\infty}dh\int_0^{\infty}\int_0^{\infty}\overline{|F(\omega,r,h,q)|^2} p(r,h,q)drdq,\nonumber\\
I_1(\omega)&=&\int_{-\infty}^{\infty}dh\int_0^{\infty}\int_0^{\infty} e^{-i\omega (r+q)}p(r,h,q)drdq,\nonumber\\
I_2(\omega)&=&\int_{-\infty}^{\infty}dh\int_0^{\infty}\int_0^{\infty} \overline{F}(\omega,r,h,q)p(r,h,q)drdq,\nonumber\\
I_3(\omega)&=&\int_{-\infty}^{\infty}dh\int_0^{\infty}\int_0^{\infty} \overline{F}^*(\omega,r,h,q)e^{-i\omega (r+q)}p(r,h,q)\nonumber\\
&&drdq,
\label{pow_3} 
\end{eqnarray}
Doing the sums in $R_K$ yields
\begin{eqnarray}
R_{K}&=&
\left\{\begin{array}{ll}
\frac{|A|^2[1-(AI_1)^{K-1}]}{(1-|A|^2)(1-AI_1)}-\frac{|A|^4(|A|^{2K-2}-(AI_1)^{K-1})}{(1-|A|^2)(|A|^2-AI_1)}, \forall \,|A|<1\\
                   \frac{K}{1-AI_1}-\frac{1-(AI_1)^K}{(1-AI_1)^2}, \mbox{for} |A|=1
              \end{array}
       \right.
\label{pow_2n} 
\end{eqnarray}
Defining $q^{\prime}=r+q$ and
$p_0(q^{\prime})=\int_{-\infty}^{\infty}dh\int_0^{\infty}p(r,h,q^{\prime}-r)dr$,
$I_1$ can be written as $I_1=\int_0^{\infty}e^{-i\omega
q^{\prime}}p_0(q^{\prime})dq^{\prime}$. $|I_1|<\int_0^{\infty}|e^{-i\omega
q^{\prime}}|p_0(q^{\prime})dq^{\prime}=1$ unless
$p_0(q^{\prime})=\sum_m
a_m\delta\left(q^{\prime}-q^{\prime}_m\right)$ where
$q^{\prime}_j-q^{\prime}_i=2\pi \ell/\omega$, ($i$, $j$ and $\ell$ are integers), $a_m> 0$ and $\sum_m a_m=1$
\cite{Mazzetti_ieeeIT67}. Thus when $K \rightarrow
\infty$, the power spectrum is ($\omega> 0$):
\begin{equation} 
\langle S_X(\omega)\rangle=
\frac{2C}{\langle r+q\rangle}\Big(I_0+2\mbox{Re}\big[\frac{AI_2I_3}{1-AI_1}\big]\Big), 
\label{pow_5} 
\end{equation}
where $C=|A|^2/\left[K(1-|A|^2)\right]+o(K^{-1})$ for all $|A|<1$, and
$C=1$ when $|A|^2=1$. 

\begin{figure}
\epsfysize=0.93\columnwidth{\rotatebox{-90}{\epsfbox{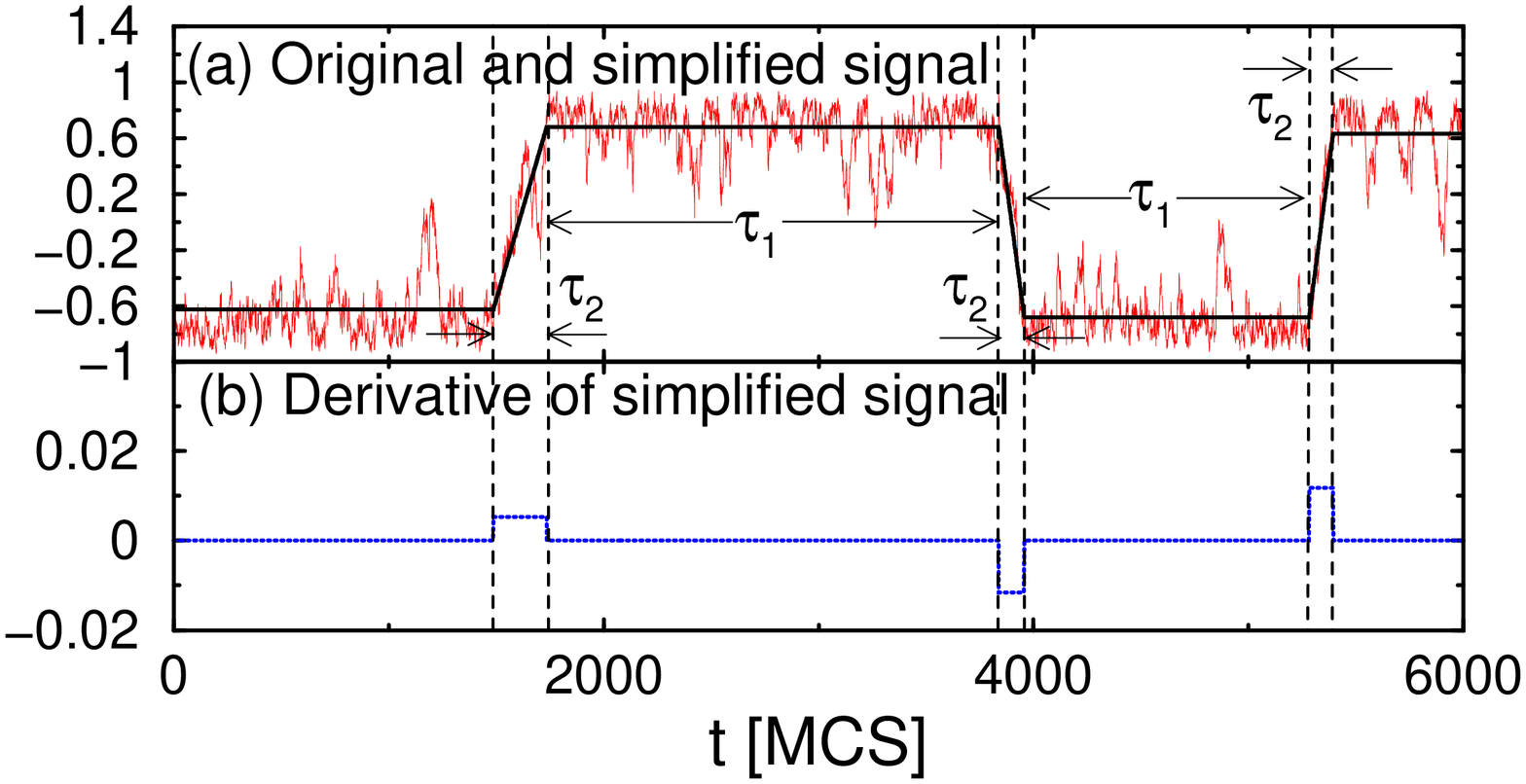}}}
\epsfysize=0.9\columnwidth{\rotatebox{-90}{\epsfbox{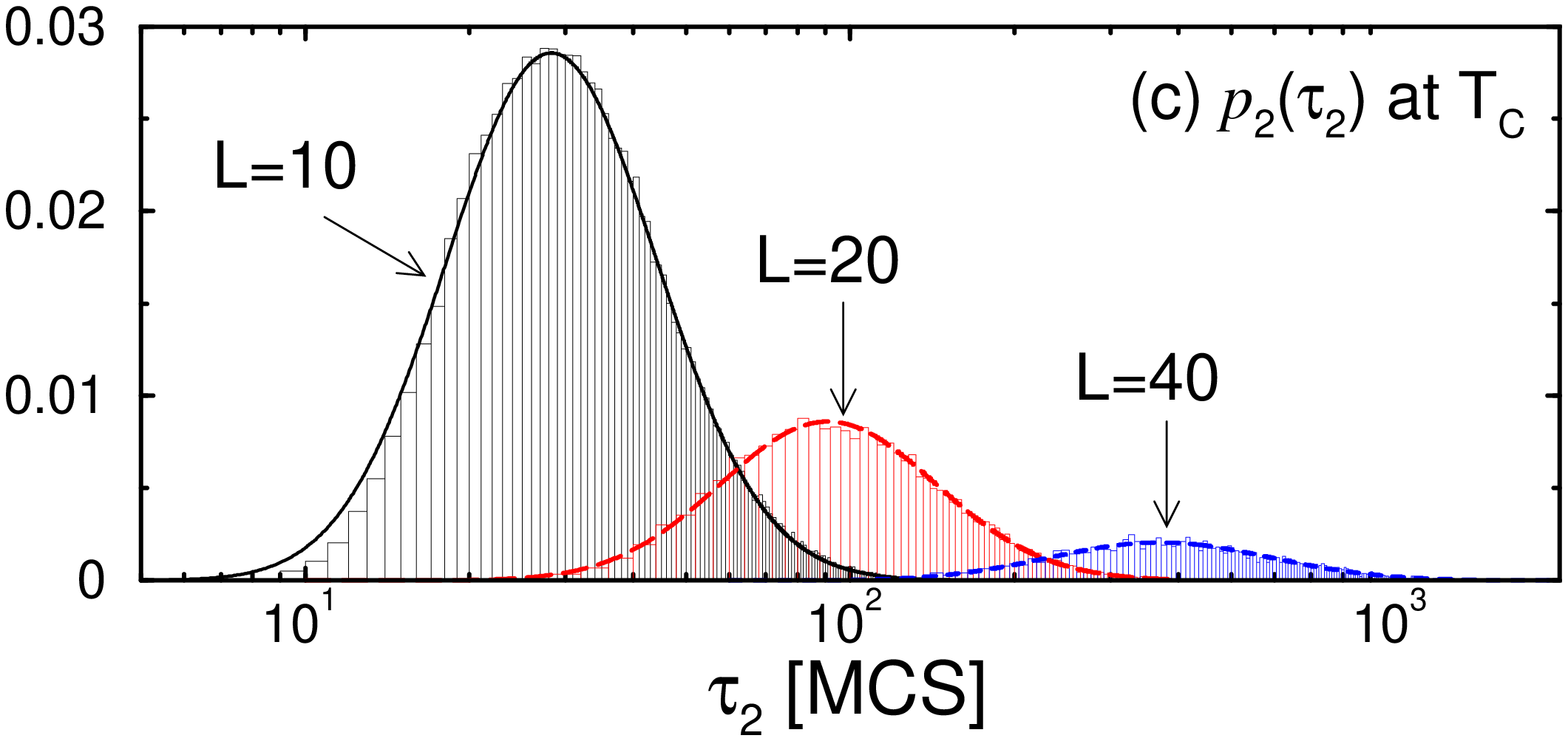}}}
\caption{(Color online) (a) The original and simplified time series
(solid lines) of the magnetization for the 2D Ising model ($L=20$) at
$T_c$. (b) The time derivative of the simplified signal shown in (a). (c)
Distributions of $\tau_2$ from simulations at $T_c$. 
Enveloping fits are $p_2(\tau_2)$ with $\langle\tau_2\rangle$ 
= 38.45, 127.3, and 517.5, $\sigma^{2}_{\tau_2}$ = 342.5, 3942.7,
and 68414, for $L$ = 10, 20, and 40, respectively.}
\label{fig:sign}
\end{figure}

\begin{figure}
\epsfysize=0.98\columnwidth{\rotatebox{-90}{\epsfbox{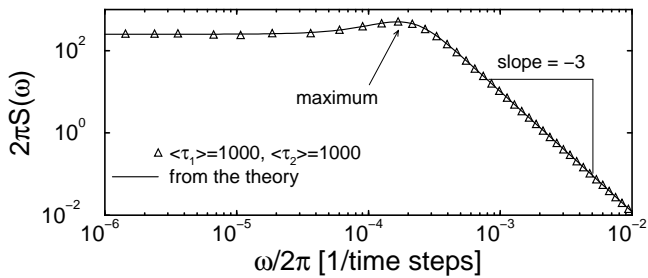}}}
\caption{Power spectra of a
trapezoidal signal ($\langle \tau_1\rangle=1000$, $\langle
\tau_2\rangle=1000$). The signal varies between 0 and 1. 
$\tau_1$ and $\tau_2$ are exponentially distributed.  Symbols are for
spectra averaged over 100 runs, each run having $2^{23}$ time
steps. The solid lines are from Eq.~(\ref{pow_5}) as described in the
text.}
\label{fig:trapezoid}
\end{figure}

Eq.~(\ref{pow_5}) can be used to calculate the average
noise spectrum of any random pulse sequence 
satisfying our three assumptions.
(Fluctuations uncorrelated with the pulses are not contained in 
Eq.~(\ref{pow_5}).) A pulse spectrum need not be a Lorentzian, e.g., it 
may have a bump. An example is a time series of 
trapezoidal pulses such as the simplified signal in Fig.~\ref{fig:sign}(a).  
This illustrates how the duration $\tau_1$ of the flat part of
the signal and the rise time $\tau_2$ are reflected in the spectra.
We find that simulated time series produces a spectrum that
agrees with the spectrum derived from Eq.~(\ref{pow_5}) as shown in
Fig.~\ref{fig:trapezoid}. The pulses are
trapezoidal when $\tau_1$ and $\tau_2$ both have an
exponential distribution. We take the time derivative of the
time series to obtain a sequence of square pulses as shown in
Fig.~\ref{fig:sign}(b) with the exponential distributions
$p_1(q)=p_1(\tau_1)$ and $p_1(r)=p_1(\tau_2)$ where $p_1(y)= \exp\left[-y/\langle
y\rangle\right]/\langle y\rangle$, $\langle y\rangle$ is the mean
value of $y$.  We use
Eq.~(\ref{pow_5}) to derive the spectrum using $A=-1$, the Fourier
transform $\overline{F}(\omega,r,h,q)$ of a square pulse, and
$p(r,h,q)=p_1(q)p_1(r)\delta(h-(1/r))$. We divide the result by
$\omega^2$ to undo the derivative.  When
$\langle\tau_1\rangle=\langle\tau_2\rangle$, Fig.~\ref{fig:trapezoid}
shows that the spectrum is flat at low frequencies with a small bump
at higher frequencies. It is well known that summing over Lorentzians
with different characteristic frequencies yields $1/f$ noise.
However Fig.~\ref{fig:trapezoid} implies that doing a similar sum
over the spectra of trapezoidal pulse sequences with
nonzero $\tau_2$ will give $1/f^{\alpha}$ noise where $\alpha > 1$.

As a physical example we now apply Eq.~(\ref{pow_5}) to study the
noise spectra of the
magnetization $M$ per spin of the 2D Ising model near the phase
transition. The Hamiltonian is:
$\mathcal{H}=-J\sum_{i<j}s_is_j$, where the spin $s_i=\pm1$, $(i,j)$
denotes the nearest neighbor sites on a square lattice and we set the
ferromagnetic exchange $J=1$. In the thermodynamic limit this model
has a second-order phase transition at the temperature
$T_c=2.2692/k_B$ \cite{Huangbook} where $k_B$ is Boltzmann's constant.

We do Metropolis Monte Carlo simulations to obtain the magnetization
time series. We apply
periodic boundary conditions to a lattice with $N=L^2$ points, where
$L$=10, 20, 30, 40. We start each run from a high
temperature ($T>T_c$), and then gradually cool the system to
$T=0.5<T_c$. We define $T_c$ for the finite system as the temperature at which
the specific heat $C_V$ has a maximum. At each temperature, we wait
until the system equilibrates (according to the protocol in
\cite{zhi_yuPRL07}) before recording the
time series for at least $10^6$ Monte Carlo time steps per spin
(MCS). The noise spectral density $S_x(\omega)$ of a time series $x(t)$ with
duration $\tau_x$ is normalized so that the total noise power per time step is
$S_{tot}=(1/\tau_x)\sum_{\omega=0}^{\omega_{max}}S_x(\omega)=\sigma^2_x$,
where $\sigma^2_x$ is the variance of $x(t)$~\cite{zhi_yuPRL07}. 

\begin{figure}
\epsfysize=0.98\columnwidth{\rotatebox{-90}{\epsfbox{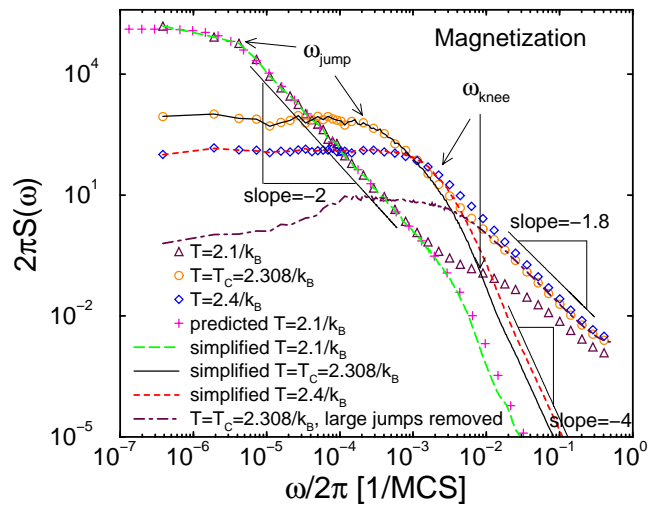}}}
\caption{(Color online) Power spectra of the original (symbols) and
simplified (lines) magnetization time series (length $N_K=2,621,440$,
10 runs) of the 2D Ising model ($L=20$) at different temperatures.}
\label{Fpow1}
\end{figure}

In Fig.~\ref{fig:sign}(a), we show an example of the magnetization time
series at $T_c$. The noise spectra are shown in Fig.~\ref{Fpow1}.
For $T\ge T_c$, starting from high frequencies, we find that the spectra
are increasing for decreasing frequencies. However, near a
characteristic frequency $\omega_{knee}$, the spectra stop
increasing, and at lower frequencies they plateau.

When cooling the system below $T_c$,
the system goes from a disordered paramagnetic phase
to an ordered ferromagnetic phase. By spin symmetry,
the distribution $P(M)$ of the magnetization per spin
is symmetric about $M=0$. So in the ordered phase the system is
equally likely to be in either an $M={\tilde M}$ or $M=-{\tilde M}$
state, where $\pm{\tilde M}$ are the most probable values of $M$. Thus for a
finite size system the magnetization may change from one state to the
other and as a result, a jump in the magnetization is created. This is
reflected in the noise spectrum. Below $T_c$ the spectrum is similar 
to the one at $T_c$ for $\omega\ge \omega_{knee}$. However, as the 
frequency decreases below $\omega_{knee}$ there is a second
increase in the spectrum which, as we shall
show, is due to magnetization flips of almost
the entire system. In this frequency range 
($\omega_{jump}<\omega<\omega_{knee}$), $S_M(\omega)\sim \omega^{-2}$
(see Fig.~\ref{Fpow1}).
Below the characteristic frequency $\omega_{jump}$, the
spectrum plateaus.

We can use our algorithm to predict the contribution of the magnetization flips
to the noise spectrum. We can approximate the time series of these jumps by a
trapezoidal signal as shown in Fig.~\ref{fig:sign}a. $\tau_1$ is the dwell
time in the flat region of up or down magnetization, and $\tau_2$ is the
duration of the jump. To apply Eq.~(\ref{pow_5}) to obtain noise spectra, 
we take the time derivative of the simplified signal to obtain a 
sequence of stochastic square pulses (see Fig.~\ref{fig:sign}b). 
Thus $\tau_1$ and $\tau_2$ correspond to $q$ and $r$ in Eq.~(\ref{pow_5}).
If each time interval has a constant probability to switch, then 
$\tau_1$ follows an exponential distribution $p_1(\tau_1)$
with mean $\langle \tau_1\rangle$: $p_1(\tau_1)=p_1(q)$.
From the noise spectrum at $T=2.1 < T_c$, we estimate 
$\langle \tau_1\rangle \sim 1/\omega_{jump}\sim 10^{5}$ MCS.
When the magnetization of the system flips, it overcomes an energy
barrier centered at $M=0$ resulting in a bimodal distribution $P(M)$.
To find $p_2(\tau_2)$, we write $\tau_2=t^a+t^b$, where $t^a$
is the time to go from $M=-{\tilde M}$ to $M=0$, and $t^b$ is the
time to go from $M=0$ to $M={\tilde M}$. We assume that $t^{a}$ and $t^{b}$
follow the same log-normal distribution
$p_{2}^{\prime}(z)=\exp[-(\ln z-\rho)^2/(2\sigma^2)]/(z\sqrt{2\pi\sigma^2})$,
where $z=t^{a}$ or $t^{b}$.
$\rho$ and $\sigma$ are determined from the mean of $\tau_2$ given by
$\langle \tau_2\rangle=2\exp[\rho+\left(\sigma^2/2\right)]$ and
the variance of $\tau_2$ given by
$\sigma_{\tau_2}^2=2(\exp[2\rho+2\sigma^2]-\langle \tau_2\rangle^2)$.
Since $\tau_2=r=t^a+t^b$, the distribution $p_2(r)=p_2(\tau_2)$ is given
by the convolution
$p_2(\tau_2)=\int dy p_2^{\prime}(y) p_2^{\prime}(\tau_2-y)$.
From the $T=2.1$ noise spectrum, we estimate 
$\langle \tau_2\rangle\sim 1/\omega_{knee}\sim 100$, and
$\sigma_{\tau_2}\sim \langle \tau_2\rangle/2\sim 50$.

Now we can apply Eq.~(\ref{pow_5}) to analyze the magnetization of the
2D Ising model. For the derivative (Fig.~\ref{fig:sign}b) of the simplified
magnetization series, $A=-1$, the Fourier transform of a
square pulse is $\overline{F}(\omega,r,h,q)=h(1-e^{-i\omega r})/(i\omega)$,
and $q$ is independent of $r$. We assume $h=1.3/\tau_2$. 
Thus the combined distribution becomes
$p(r,h,q)=p_1(q)p_2(r)\delta(h-1.3/\tau_2)$.
Doing the integrals in Eq.~(\ref{pow_3}) numerically, and dividing the
result of Eq.~(\ref{pow_5}) by $\omega^2$ to undo the derivative, we
obtain the spectra shown in Fig.~\ref{Fpow1} which is a good approximation
at low frequencies. Thus we predict that pulses dominate the low frequency 
spectrum.

We can check this prediction by extracting a simplified trapezoidal signal 
that represents a series of magnetization jumps as shown in Fig.~\ref{fig:sign}a.
In the flat regions with
small fluctuations, we replace the original magnetization time
series by the mean magnetization in that region. To find the jumps, we
start from points with $M=0$, then move both forward and backward in
time. A jump is identified if and only if $M$ in one direction
achieves ${\tilde M}-\delta M$, and in the other direction achieves
$-{\tilde M}+\delta M$, where $\delta M > 0$ is the offset. 
In Fig.~\ref{Fpow1}, for $T<T_c$, we find
that the spectra of the original and simplified magnetization series
match at low frequencies ($\omega\ll \omega_{knee}$).
At $T_c$ up to frequencies one order of
magnitude higher than the crossover frequency, the power spectrum for
the simplified signal fits the spectrum of the original signal very
well. This is why this contribution must be subtracted from the spectrum 
before extracting the high frequency power law dictated by the
critical exponents. In Fig.~\ref{Fpow1} the high frequency noise at
$T_c$ follows $S(\omega>\omega_{knee})\sim
\omega^{-\mu_M}$ where $\mu_M = 1.8$. This matches well with the scaling
theory prediction \cite{zhi_yuPRL07}: $\mu_M=1.8$.

\begin{figure}
\epsfysize=0.98\columnwidth{\rotatebox{-90}{\epsfbox{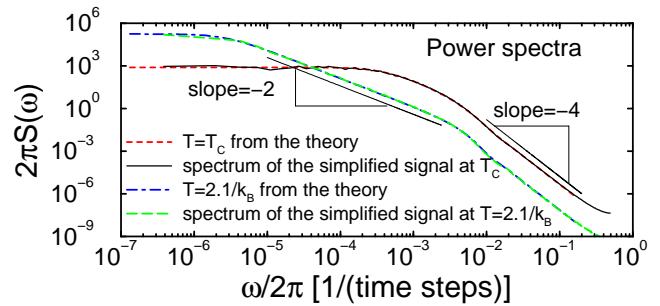}}}
\caption{(Color online) Power spectra for the magnetization jumps of
the simplified magnetization signal at $T=T_c$ and $T=(2.1/k_B)<T_c$
for $L=20$.  At $T_c$, $\langle \tau_1 \rangle=1030.7$, $\langle
\tau_2 \rangle=127.3$, $\sigma_{\tau_2}^2=3942.7$.  At $T=2.1/k_B$,
$\langle \tau_1 \rangle=1.169\cdot 10^5$, $\langle \tau_2
\rangle=115.1$, $\sigma_{\tau_2}^2=1718.3$.}
\label{pow_simu}
\end{figure}

From the simplified time series, we can extract the parameters for
the pulse distributions to derive the pulse contribution
to the noise by using Eq.~(\ref{pow_5}) in the same
way as we did for the estimate. We use
$h=C(T)/\tau_2$ where $C(T)$ is a function of the temperature $T$.
We use the same distributions for $\tau_1$ and $\tau_2$, and obtain
values for $\langle \tau_1\rangle$, $\langle\tau_2\rangle$,
and $\sigma_{\tau_2}^2$ from the simplified time series.
As shown in Fig.~\ref{fig:sign}(c), $p_2(\tau_2)$
fits very well with the
actual distribution of $\tau_2$ for different system sizes.
We obtain the spectra shown in Fig.~\ref{pow_simu} which
is an excellent fit to the spectra of the simplified magnetization signals
for $T\leq T_c$.  The square-wave-like time series of the derivative
of the simplified signal (Fig.~\ref{fig:sign}b) yields a Lorentzian-like power
spectrum. Undoing the derivative with a factor of $1/\omega^2$ yields a high
frequency noise spectrum that goes as $1/\omega^4$.

In summary, our method predicts and derives the noise spectra
of stochastic pulse sequences from the distribution of pulse 
parameters. It can be used to separate out the contribution
of pulses from other contributions to the spectra. It can even
be used for noise spectra not derived from a time series.

This work was supported by DOE grant DE-FG02-04ER46107.

\end{document}